\documentclass[aps,preprint]{revtex4}

\usepackage{psfrag}
\usepackage{graphicx}

\usepackage{amssymb}
\usepackage{bm}
\begin{document}
\setcounter{page}{1}
\preprint{RH-13-2006}
\title[]{Cosmic censorship inside black holes}
\author{L\'arus \surname{Thorlacius}}
\email{lth@hi.is}

\affiliation{Science Institute, University of Iceland, Dunhaga 3,
107 Reykjavík, Iceland\\ \ }

\begin{abstract}
A simple argument is given that a traversable Cauchy 
horizon inside a black hole is incompatible with unitary 
black hole evolution. The argument assumes the validity 
of black hole complementarity and applies to a generic 
black hole carrying angular momentum and/or charge. 
In the second part of the paper we review recent work 
on the semiclassical geometry of two-dimensional charged 
black holes. 
\end{abstract}

\maketitle

\section{INTRODUCTION}
A stationary rotating black hole in classical general relativity 
is described by the maximally extended Kerr spacetime, which has 
multiple asymptotic regions and black hole regions that contain
timelike singularities. For a massive black hole with large angular 
momentum there exist timelike paths connecting different asymptotic 
regions without ever entering regions of strong curvature and one 
can speculate whether macroscopic observers could travel along 
such paths through the black hole to another ``universe". 
There are also paths in the Kerr geometry that lead past a 
singularity into into an exotic region with closed timelike curves. 
The sperically symmetric Reissner-Nordstr\"om spacetime of a static 
electrically charged black hole also has multiple asymptotic 
regions and timelike singularities but no closed timelike curves.
The Penrose diagram of maximally extended Reissner-Nordsr\"om
spacetime is shown in Figure~1 and, for comparison, the Penrose
diagram for an uncharged Schwarzschild black hole is shown in 
Figure~2.

\begin{figure}[t!]
\includegraphics[width=3.5cm]{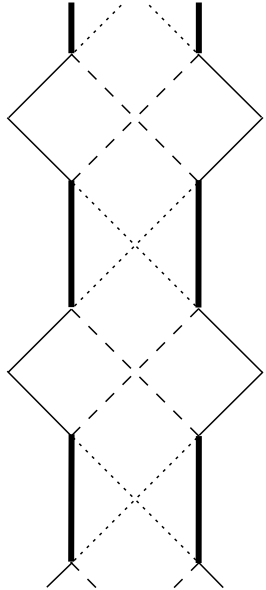}
\caption{Penrose diagram for maximally extended 
Reissner-Nordstr\"om spacetime. Thick lines represent timelike
singularities, dashed lines are event horizons, and dotted lines 
are Cauchy horizons. The diagram is periodic in the vertical 
direction with an infinite number of asymptotic regions.} 
\label{fig.1}
\end{figure}

\newpage
\begin{figure}[t!]
\includegraphics[width=5.0cm]{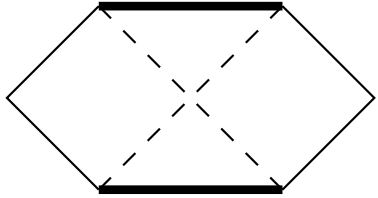}
\caption{Penrose diagram for maximally extended
Schwarzschild spacetime. Thick lines represent spacelike
singularities and dashed lines are event horizons.}
\label{fig.1a}
\end{figure}

In both the Kerr and Reissner-Nordstr\"om cases the physical 
relevance of much of the extended spacetime has, however, been 
called into question \cite{Penrose:1968,Simpson:1973,
McNamara:1978,Novikov:1980ni,Chandrasekhar:1982,Zamorano:1982}. 
Inside each black hole region there is a Cauchy horizon, 
beyond which a timelike singularity becomes visible. 
The Cauchy horizon is a surface of infinite blueshift with 
respect to the event horizon of the black hole in question and 
this leads to a dynamical instability, referred to as mass 
inflation, which replaces the Cauchy horizon by a null 
singularity that turns spacelike deep inside the black 
hole \cite{Poisson:1990eh,Ori:1991,Brady:1995ni,Hod:1998gy,
Dafermos:2003wr}. However, the null singularity is weak in the 
sense that integrated tidal effects acting on extended observers 
are finite and this leaves open the possibility of extending the 
physical spacetime through it \cite{Ori:1991}. 
Such an extension requires ingoing flux of negative energy along
the null singularity itself and while negative energy is 
pathological at the classical level this not necessarily the case 
in the semiclassical context of quantum field theory in a curved 
background spacetime. 

In view of this it is natural to ask how quantum effects modify 
the classical geometry inside a rotating or charged black hole. 
Do they render the Cauchy horizon traversable by supplying 
the flux of negative energy that is required for extending the
geometry? Or, is the Cauchy horizon further destabilized by
quantum effects and replaced by a spacelike 
singularity \cite{Novikov:1980ni,Herman:1994nv}?
In the following we present two very different approaches to this
problem that both support the latter possibility. The first one,
discussed in Section~II, is based on Hawking's black hole 
information paradox \cite{Hawking:1976ra}. It is qualitative and 
rather speculative but applies to generic black holes, with or 
without angular momentum and charge. 
The other approach is quantitative but more restricted in 
scope. It involves charged black holes in a two-dimensional
toy model of gravity where quantum effects can be studied
systematically \cite{Frolov:2005ps,Frolov:2006is}. Section~III 
contains a brief review of recent work on this model.

\section{A perspective from black hole complementarity}
In this section we argue that a traversable Cauchy horizon 
inside a black hole formed in gravitational collapse is 
incompatible with quantum mechanical unitarity. The argument 
assumes the validity of the principle of black hole 
complementarity \cite{Susskind:1993if} but is otherwise quite 
general and applies to both rotating and charged black holes.

\subsection{Unitary black hole evolution}
Let us adopt the point of view that the formation and subsequent 
evaporation of a rotating and/or charged black hole is a unitary 
process \cite{Page:1979tc,tHooft:1991rb}. A number of theoretical 
developments, including matrix theory \cite{Banks:1996vh}, the 
gauge theory/gravity correspondence \cite{Maldacena:1997re}, 
and the microscopic computation of black hole entropy in string 
theory \cite{Strominger:1996sh}, support this viewpoint even if 
the detailed implementation of unitarity has yet to be understood 
(see \cite{Lowe:2006xm,Giddings:2006sj} for recent discussions of 
unitarity and the black hole information paradox).

Let us further assume that unitarity is maintained by encoding 
the information about the initial quantum state of the matter 
that forms the black hole into subtle correlations in the 
outgoing Hawking radiation. In principle, this information
eventually becomes available to observers outside the black
hole although quite hard to come by in practise. The late time
Hawking radiation will also contain information about the state
of an observer that falls into the black hole during its lifetime
but this appears to contradict the expectation, based on the
equivalence principle, that an observer in free fall encounters 
nothing unusual upon crossing the event horizon of a large
black hole. 

Quantum states cannot be cloned \cite{Wooters:1982,Dieks:1982} 
and therefore, as
far as distant observers are concerned, any information that 
emerges in the outgoing Hawking radiation must be removed 
from the infalling matter before it enters the black hole.
On the other hand, from the point of view of an infalling 
observer who passes unharmed through the event horizon no 
information has been removed and its duplication in the 
Hawking radiation appears to violate the principles of quantum 
mechanics.  

\subsection{Gedanken experiment involving correlated spins}
The principle of black hole complementarity states that both 
viewpoints are equally valid and that apparent contradiction 
only comes about when we attempt to compare the physical 
description in the very different reference frames of these 
observers. In \cite{Susskind:1993mu} this was illustrated by
analysing various gedanken experiments involving black holes 
designed to expose potential contradictions. 

\begin{figure}[t!]
\psfrag{o1}{${\cal O}_1$}
\psfrag{o2}{${\cal O}_2$}
\includegraphics[width=6.0cm]{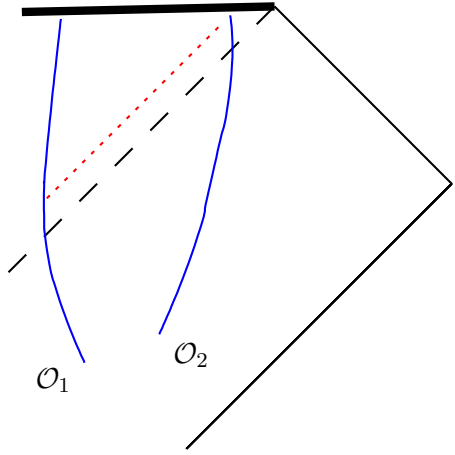}
\caption{Gedanken experiment for a black hole with a 
spacelike singularity (thick line). The dashed line is the 
event horizon and the dotted line represents a null signal from 
${\cal O}_1$ which is to be received by ${\cal O}_2$ before 
${\cal O}_2$ hits the singularity.} 
\label{fig.2}
\end{figure}

Let us consider one of these experiments which involves
quantum correlations between degrees of freedom inside and
outside the event horizon. A pair of spins is prepared in a 
singlet state outside a large black hole and then one of the 
spins is carried into the black hole by observer ${\cal O}_1$
in Figure~\ref{fig.2} while the other spin remains outside
the black hole and is never actually measured in the 
experiment. Here ``spin" is taken to mean some internal label 
because conventional spin can in principle be detected at 
long-range by its gravitational field. 

Upon crossing the event horizon, which is indicated by a 
dashed line in the Figure~\ref{fig.2}, the infalling observer 
${\cal O}_1$ carries out a measurement of the spin and 
transmits the result via a null signal, shown as a dotted line 
in the figure.  Meanwhile observer ${\cal O}_2$ makes 
measurements on the Hawking radiation coming from the 
black hole. Assuming all information about the quantum state 
inside the black hole is encoded in the Hawking radiation
${\cal O}_2$ can effectively measure a component of the 
spin that went inside the black hole. ${\cal O}_2$ then 
passes inside the event horizon where he can receive 
the signal from ${\cal O}_1$, which potentially contradicts
the measurement of the Hawking radiation, in violation
of the laws of quantum mechanics. 

If the black hole singularity is spacelike, as in 
Figure~\ref{fig.2}, there is no paradox here. The information
in the Hawking radiation is not immediately available.
In fact, less than one bit of information is accessible until 
the black hole area has evaporated to half its initial value
and after that it comes out gradually \cite{Page:1993df}.
Observer ${\cal O}_2$ must therefore wait outside the 
evaporating black hole for a time, which is of order the
black hole lifetime, before the measurements of the 
Hawking radiation can yield any useful information. 
By then the extreme redshift of the region close to the 
event horizon, prevents ${\cal O}_2$ from learning of 
the potential contradiction before hitting the singularity. 
The proper time available for ${\cal O}_1$ to carry out 
the spin measurement and transmit the signal, so that it 
can be received by ${\cal O}_2$, turns out to be much 
shorter than a Planck time \cite{Susskind:1993mu}.  

\begin{figure}[t!]
\psfrag{o1}{${\cal O}_1$}
\psfrag{o2}{${\cal O}_2$}
\includegraphics[width=6.0cm]{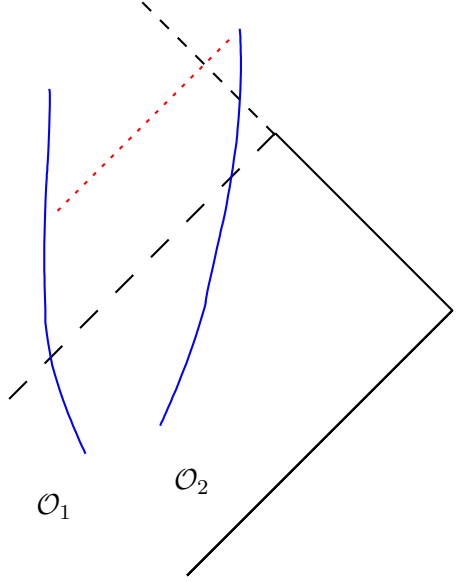}
\caption{Gedanken experiment for a black hole with a
traversible Cauchy horizon. In this case ${\cal O}_2$ can receive
signals after passing through the Cauhy horizon.}
\label{fig.3}
\end{figure}

If, on the other hand, the black hole has a traversable
Cauchy horizon, as in Figure~\ref{fig.3}, then ${\cal O}_2$
is at leisure to receive signals after passing through the
Cauchy horizon and there is no stringent time limit for 
${\cal O}_1$ to carry out the spin measurement and 
transmit the result. A black hole with a traversable 
Cauchy horizon is therefore incompatible with black
hole complementarity and we conclude that black holes
cannot have traversable Cauchy horizons. 

We note that the same gedanken experiment was considered 
in \cite{Ge:2005bn} in a Reissner-Nordstr\"om background 
geometry. Those authors did not question the validity of 
the classical static geometry but instead concluded that 
the experiment presents a problem for unitarity and the 
principle of black hole complementarity.

\section{Two-dimensional charged black holes}
In this section we review our recent work with A.~Frolov and
K.~Kristjansson \cite{Frolov:2005ps,Frolov:2006is} where
we study the semiclassical geometry of electrically charged
black holes in the simplified context of 1+1-dimensional
dilaton gravity coupled to an abelian gauge field. 
The model has classical charged black hole solutions 
for which the maximally extended spacetime has 
multiple asymptotic regions separated by black hole 
regions containing timelike singularities and
associated Cauchy horizons. The Penrose diagram is 
in fact identical to that of a 3+1-dimensional 
Reissner-Nordstr\"om black hole, shown in Figure~\ref{fig.1}. 
Furthermore, mass inflation has been shown to occur at the 
classical level in 1+1-dimensional dilaton gravity models of 
this type \cite{Balbinot:1994ee,Chan:1994tb,Droz:1994aj},
suggesting that this is indeed a suitable toy model to study 
the fate of Cauchy horizons at the quantum level. 

The main new feature of \cite{Frolov:2005ps,Frolov:2006is} 
is to include quantum effects due to electrically charged
matter, which turns out to significantly modify the 
internal geometry of charged black holes in the model.
The timelike singularities and Cauchy horizons of a 
static classical black hole are replaced by a spacelike
singularity at the semiclassical level and the global 
topology becomes the same as that of an electrically
neutral Schwarzschild black hole. This conclusion is
reached by a combination of analytic and numerical
calculations that we only outline here, referring to 
\cite{Frolov:2005ps,Frolov:2006is} for a more detailed
description. 

\subsection{Classical theory}
Our starting point is the classical action,
\begin{equation}
\label{classact}
S_ {0}  = \int d^2x \sqrt{-g}e^{-2\phi}
\left[R+4(\nabla\phi)^2+4\lambda^2-\frac{1}{4} F^2\right],
\end{equation}
with a dilaton field $\phi$, an abelian gauge field 
$F_{\mu\nu}$ and the 1+1-dimensional metric 
$g_{\mu\nu}$. Due to the factor of $e^{-2\phi}$ in front,
the strength of both the gravitational and gauge couplings
is determined by the value of the dilaton field. 

The model can be viewed as the s-wave reduction of 
3+1-dimensional dilaton gravity in an extremal 
black hole background \cite{Callan:1992rs,Giddings:1992kn,
Banks:1992ba}. The mass scale $\lambda$ is proportional
to the inverse of the magnetic charge carried by the
3+1-dimensional extremal dilaton black hole. 
We will use units where $\lambda =1$. We also note that 
the two-dimensional dilaton field $\phi$ has an interpretation
in terms of an area in the higher dimensional theory. The
precise relation is that the area of the transverse two-sphere
in the Einstein frame of 3+1-dimensional dilaton gravity
is given by $\psi \equiv e^{-2\phi}$ and thus we
refer to $\psi$ as the {\it area function}. 

The action (\ref{classact}) has classical solutions, which are 
analogous to Reissner-Nordstr\"om black holes,
\begin{eqnarray}
\label{bhone}
\phi  &=& -x , \\
\label{lindilmetric}
ds^2
&=& -a(x)dt^2+\frac{1}{a(x)}dx^2  ,\\
\label{ffield}
F_{tx}  &=& Q e^{-2x} ,
\end{eqnarray}
with
\begin{equation}
\label{afunction}
a(x)=1-Me^{-2x}+\frac{1}{8}Q^2e^{-4x}.
\end{equation} 
The constants $M$ and $Q$ are the mass and charge and
for $M>\vert Q\vert/\sqrt{2}$ we have a classical black hole 
geometry with horizons located at the two zeroes of the 
metric $a(x)$, where the area function takes the 
following
values,
\begin{equation}
\psi_\pm=\frac{1}{2}\left(M\pm \sqrt{M^2-\frac{1}{2}Q^2} \right) .
\label{psipm}
\end{equation}
The coupling strength $e^\phi$ goes to zero in the asymptotic 
region $x\rightarrow\infty$ and the metric approaches the
1+1-dimensional Minkowski metric there. On the other hand, 
there is a curvature singularity inside the black hole, where 
the area function goes to zero and the coupling diverges.

\subsection{Charged matter}
We now add matter to the model to be able to consider 
dynamical solutions involving gravitational collapse and
also to study semiclassical corrections to the geometry
due to matter quantum effects. In order to form charged 
black holes by gravitational collapse we need to have some
form of electrically charged matter and we take this to be
charged Dirac fermions,
\begin{equation}
\label{matteract}
S_ {m}  = \int d^2x \sqrt{-g}\left[
i\bar\Psi\gamma^\mu ({\cal D}_\mu+ieA_\mu )\Psi
- m\bar\Psi\Psi \right]
\end{equation}
where $e$ and $m$ are the fermion charge and mass respectively,
and $A_\mu$ is the 1+1-dimensional gauge potential.

At the quantum level, fermions will be pair-produced in 
the electric field of a charged black hole via the Schwinger
effect \cite{Schwinger:1951} and we are particularly 
interested in the resulting electromagnetic screening of
the black hole charge and the gravitational backreaction 
on the black hole geometry due to the charged pairs. We are 
also interested in quantum effects due to electrically neutral 
matter, {\it i.e.} Hawking emission \cite{Hawking:1974sw} of 
neutral particles from our charged black holes, but let us 
first consider the effect of Schwinger pair-production and 
come back to include the Hawking effect later on.

In our two-dimensional model pair-production of charged 
particles is most conveniently described by bosonizing the
fermions. This has two important advantages. First of all,
the matter equations of motion are then converted into 
scalar field equations which are simpler to analyze both
analytically and numerically than the original fermion
equations. The second advantage is that, since bosonization
is a strong-to-weak coupling duality, the quantum mechanical
process of fermion pair-production is well described by
classical equations for the bosons. This holds when 
$m\ll e$ and that happens to be the range of parameters of
most interest to us because of the very large charge-to-mass
ratio of electrons in the real world.

The identification \cite{Coleman:1975pw,Coleman:1976uz} 
between composite operators of the fermion field and of a 
real boson field $Z$ carries over from flat to curved 
spacetime \cite{Grumiller:2006}, with appropriate replacement 
of derivatives by covariant derivatives, as long as the 
background geometry is slowly varying on the microscopic
length scale of the matter system. The bosonic description
thus breaks down near singularities where the spacetime 
curvature gets large on microscopic length scales, but
this is not a problem for the bosonization {\it per se} 
since all classical or semiclassical equations are 
inadequate in such regions. 

The electromagnetic current is given by 
\begin{equation}
\label{current}
j^\mu
= e\bar\Psi\gamma^\mu\Psi 
= \frac{e}{\sqrt{\pi}}\varepsilon^{\mu\nu}\nabla_\nu Z,
\end{equation}
where $\varepsilon^{\mu\nu}$ is an antisymmetric tensor
related to the Levi-Civita tensor density by 
$\varepsilon^{\mu\nu}=\epsilon^{\mu\nu}/\sqrt{-g}$.
The boson effective action is 
\begin{equation}
\label{bosonact}
S_ {b}  = \int d^2x \sqrt{-g}\left[-\frac{1}{2} (\nabla Z)^2
- V(Z) -\frac{e}{\sqrt{4\pi}} \varepsilon^{\mu\nu}F_{\mu\nu} 
Z\right],
\end{equation}
where $V(Z)=c\,e\,m (1-\cos(\sqrt{4\pi}Z))$ and $c$ is an 
$O(1)$ numerical constant whose precise value will not be
needed.

\subsection{Semiclassical black holes}
The semiclassical geometry of charged black holes is 
obtained by solving the equations of motion of 
the combined dilaton gravity and boson action
(\ref{classact}) and (\ref{bosonact}). We work in
conformal gauge 
$ds^2=-e^{2\rho}d\sigma^+d\sigma^-$
and take advantage of the fact that a two-dimensional
gauge field is non-dynamical. The two-dimensional 
Maxwell equations can be used to express the field
strength in terms of the dilaton and the bosonized
matter field,
\begin{equation}
\label{zforgauge}
F^{\mu\nu} = -\left(-\frac{e}{\sqrt{\pi}}Z+q \right)
                        e^{2\phi} \varepsilon^{\mu\nu},
\end{equation}
where $q$ is a constant of integration which can be
interpreted as a background charge located at the 
strong-coupling end of the one-dimensional space.
We are primarily interested in gravitational collapse
of charged matter into an initial vacuum and then
it is natural to set $q=0$. We also note that the
value of $Z/\sqrt{\pi}$ at a given location determines
the amount of electric charge to the left of, or 
``inside", that location. This can also be inferred 
from the bosonized current in (\ref{current}).
It follows that electromagnetic screening will 
appear as spatial variation in $Z$ in the bosonized
formalism.

The remaining semiclassical equations are 
simplified if we introduce $\theta=2(\rho-\phi)$
and work with the area function $\psi$ rather
than the dilaton itself. This leads to a system 
of equations,
\begin{eqnarray}
\label{psieq}
-4\partial_+\partial_-\psi
&=& \left(4-\frac{e^2Z^2}{2\pi\psi^2}\right)e^\theta
-\frac{V(Z)e^\theta}{\psi} , \\
\label{thetaeq}
-4\partial_+\partial_-\theta
&=& \frac{e^2Z^2e^\theta}{\pi\psi^3}
+\frac{V(Z)e^\theta}{\psi^2}, \\
\label{zeq}
-4\partial_+\partial_-Z
&=& \frac{e^2Ze^\theta}{\pi\psi^2} +
\frac{V'(Z)e^\theta}{\psi} ,
\end{eqnarray}
along with two constraints
\begin{equation}
\label{constraints}
\partial_\pm^2\psi-\partial_\pm\theta\partial_\pm\psi 
= -\frac{1}{2}(\partial_\pm Z)^2.
\end{equation}
In \cite{Frolov:2005ps,Frolov:2006is} this system was
solved numerically for initial data consisting of a 
left-moving kink configuration in the bosonized matter
field, corresponding to incoming charged matter 
undergoing gravitational collapse. The height of
the kink determines the total incoming charge
while its steepness governs the total incoming 
energy. 

\begin{figure}[t!]
\rotatebox{45}{
\includegraphics[width=5.0cm]{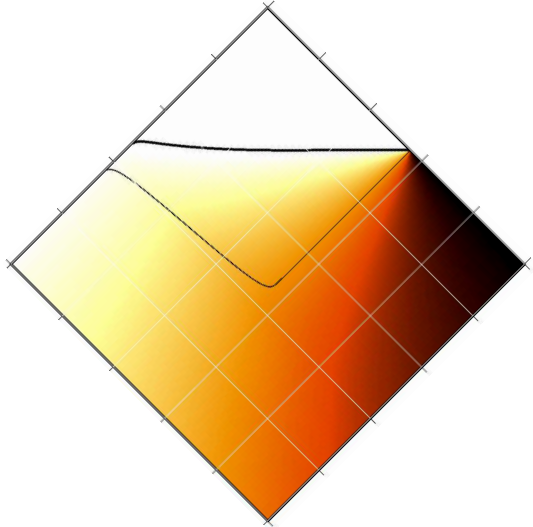}
}
\caption{Numerical black hole formation. The figure shows a 
density plot of the area function $\psi$ for an incoming
pulse of charged matter. The 1+1-dimensional spacetime is 
represented in (compactified) null coordinates. The curvature 
singularity is indicated by the thick black curve and the 
apparent horizon by the thin curve. The singularity is 
spacelike.}
\label{fig.4}
\end{figure}

\begin{figure}[t!]
\rotatebox{45}{
\includegraphics[width=5.0cm]{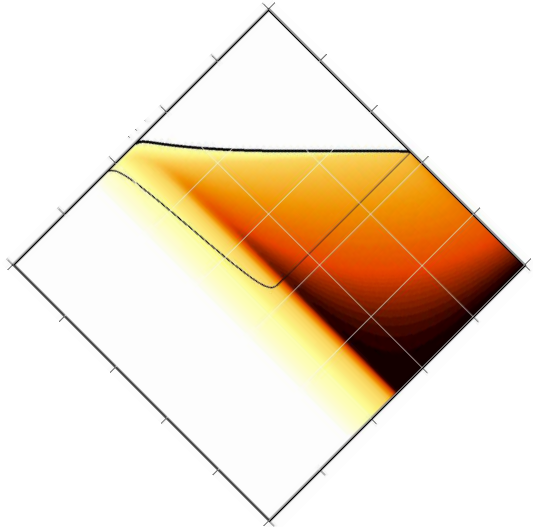}
}
\caption{Numerical evolution of an incoming pulse of 
charged matter. The density plot shows the bosonized
matter field $Z$, which represents the charge distribution.
The white region on the lower left is the vacuum before the 
arrival of the matter pulse. Charge screening that prevents 
the $Z$ field from penetrating into the strong-coupling 
region near the singularity is evident.}
\label{fig.5}
\end{figure}

Figures \ref{fig.4} and \ref{fig.5} show density 
plots for the area function $\psi$ and the 
bosonized matter field $Z$ for a typical solution.
The geometry is represented in null coordinates
and we have performed a conformal compactification
in order to cover the entire spacetime in the plots.
Curvature singularities, which occur at $\psi=0$,
are indicated by thick black curves and apparent
horizons, where $\partial_+\psi=0$ or 
$\partial_-\psi=0$ \cite{Russo:1992ht}, by thin
black curves.

It is clear from these plots that a charged black
hole forms with a curvature singularity inside an 
apparent horizon. The bosonic matter field goes to 
zero deep inside the black hole. This reflects 
charge screening due to fermion pair-production.
In fact, for this particular solution the screening
effect extends well outside the black hole region.
We also note that after the black hole has formed
the value of $Z$ decreases along the apparent
horizon, {\it i.e.} the black hole gradually discharges
and the final state is a neutral black hole.

The black hole singularity is everywhere spacelike
and approaches the apparent horizon at future
null infinity. There is no indication of a Cauchy 
horizon or a null singularity inside the semiclassical
black hole.

\subsection{Static solutions}
\begin{figure}[t!]
\begin{tabular}{c@{\hspace{1.0cm}}c}
\includegraphics[width=4cm]{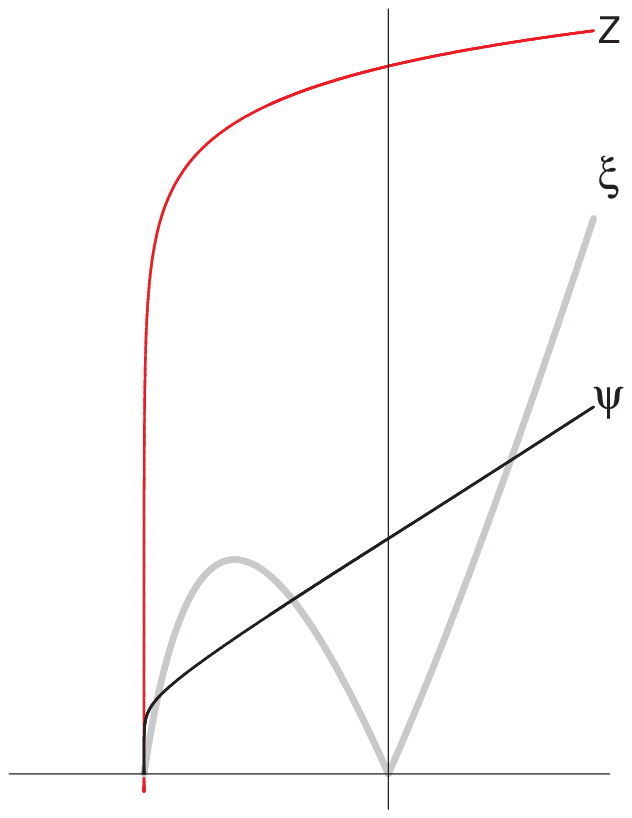} &
\includegraphics[width=4cm]{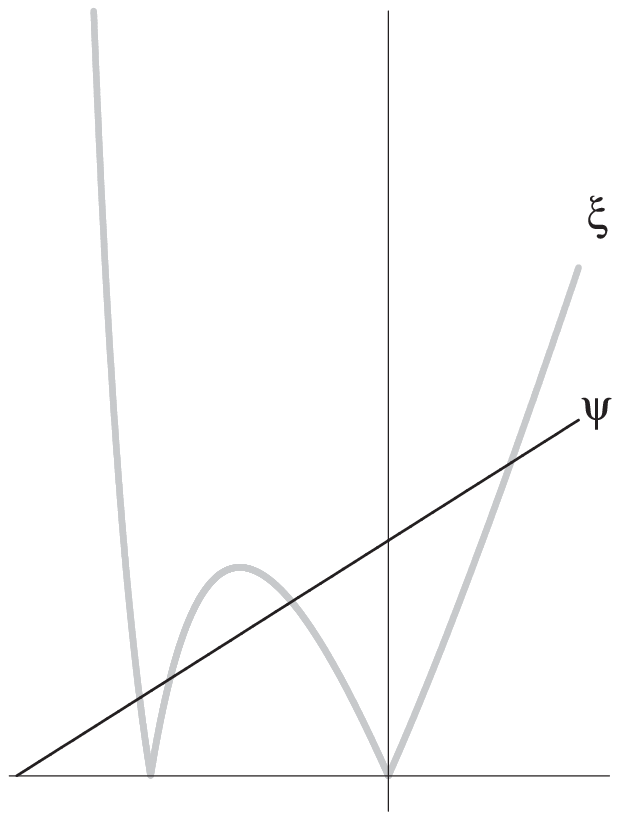} \\
(a) & (b)\\
\end{tabular}
\caption{(a) Semiclassical static black hole solution. The numerical
integration is started with initial conditions that ensure smoothness
at the event horizon, which is located at the origin of the spatial 
coordinate. The fields $Z$, $\xi$ and $\psi$ all approach zero at
a singularity inside the black hole. (b) The corresponding classical
black hole solution with a smooth inner horizon. At the classical 
level there is no charge screening so the black hole carries fixed 
charge and there is no $Z$ field.}
\label{fig.6}
\end{figure}

One can also study static black holes in this 
semiclassical theory. For this we require the
fields to only depend on a spatial variable 
$\sigma= \frac{1}{2}(\sigma^+ - \sigma^-)$
and make a further change of variables,
writing $\xi=e^\theta$ and defining a new
spatial coordinate via $dy = \xi d\sigma$.
In these variables the classical charged black
hole solution (\ref{bhone})-(\ref{afunction}) can 
be extended inside the black hole horizon in a
straightforward way and the corresponding 
extension for a semiclassical black hole can
be carried out numerically, as described in
detail in \cite{Frolov:2005ps,Frolov:2006is}. 
The result of such a numerical integration 
is shown in Figure~\ref{fig.6}a, with the 
corresponding classical solution shown in
Figure~\ref{fig.6}b for comparison. 

The classical solution has two horizons 
where $\xi(y)$ goes to zero. One is the event 
horizon and we have chosen to put the 
origin of our spatial coordinate $y=0$ at 
its location. There is also a smooth inner
horizon, the Cauchy horizon, at some 
negative value of $y$.

The numerical solution of the semiclassical
equations is started at $y=0$ and the 
initial data chosen so as to obtain a black 
hole with a smooth event horizon
\cite{Birnir:1992by}. To obtain a smooth inner 
horizon also requires some tuning of parameters
which turns out, however, to be incompatible
with the requirements for a smooth event 
horizon. In other words, we are not able to
find static semiclassical solutions with two
smooth horizons. Instead we encounter a 
singularity inside the black hole, where 
$\psi$, $\xi$ and $Z$ all go to zero. As we
approach the singularity our numerical 
solution eventually breaks down but, as far 
as the numerics go, the indication is that 
we are dealing with a spacelike curvature 
singularity. The notion of a strong spacelike 
singularity is also supported by analytic 
scaling solutions, presented in 
\cite{Frolov:2006is}, which we believe to be 
asymptotic to an exact solution sufficiently 
near the singularity. 

The Cauchy horizon of a classical static black 
hole is replaced by a spacelike singularity 
at the semiclassical level.
This can be traced to charge screening due to
pair-production inside the black hole. For a 
static semiclassical solution the constraint 
equations (\ref{constraints}) reduce to 
\begin{equation}
\frac{d^2\psi}{dy^2} = -\frac{1}{2}
\left( \frac{dZ}{dy} \right)^2 .
\end{equation}
As we go deeper into the black hole, {\it i.e.} 
towards negative $y$, any change in $Z$ will 
focus the area function towards zero faster 
than in the corresponding classical solution 
(where $\psi$ is linear in $y$). As discussed above, 
charge screening appears as spatial variation in 
$Z$ and therefore screening contributes to the
focusing of the area function. This ultimately leads 
to the singular behavior seen in Figure~\ref{fig.6}a.

\subsection{Neutral matter and the Hawking effect}
So far we have considered semiclassical effects 
involving charged matter but we can also include
electrically neutral matter along the lines of
\cite{Callan:1992rs} by introducing conformally
coupled scalar fields $f_i$, with flavor index
$N=1,\ldots,N$,
\begin{equation}
S_f= -\frac{1}{2}\int d^2x \sqrt{-g} 
\sum_{i=1}^N \left(\nabla f_i \right)^2.
\end{equation}
This gives rise to the following non-local term
in the semiclassical effective action,
\begin{equation}
\label{pterm}
-\frac{N}{48} \int d^2 x \sqrt{-g(x)} \int d^2 x' 
\sqrt{-g(x')} R(x) G(x,x') R(x') ,
\end{equation}
in a leading $1/N$ approximation for $N\gg 1$, 
where $G(x,x')$ is a Green function for the
$\nabla^2$ operator. The analysis of semiclassical
models of this type (see \cite{Giddings:1994pj,
Strominger:1994tn, Thorlacius:1994ip, 
Grumiller:2002nm,Fabbri:2005mw} for reviews) 
is usually carried out in conformal gauge 
where (\ref{pterm}) reduces to 
\begin{equation}
\label{cgaugepterm}
-2\kappa \int d^2 x \partial_+\rho \partial_-\rho,
\end{equation}
with $\kappa \equiv N/12$. In \cite{Frolov:2006is} 
the so-called RST-term \cite{Russo:1992ax,Russo:1992yh},
\begin{equation}
\label{rstterm}
-\frac{N}{48} \int d^2 x \sqrt{-g}  R \phi =
- 2\kappa \int d^2x \phi \partial_+\partial_-\rho ,
\end{equation}
was also included in the action. In that case the
model has exact semiclassical solutions for 
electrically neutral black holes which can be 
useful even if our main interest here is in 
charged black holes. 

A closely related model was studied in
\cite{Ori:2001xc}, where analytic results were
obtained for the combination of Hawking radiation 
and pair-production of charged particles in
the background geometry of a classical 
two-dimensional charged black hole, but
the semiclassical back-reaction on the geometry,
which is the main concern of \cite{Frolov:2006is},
was not considered. 

The semiclassical equations become more 
complicated when the additional terms that come 
from (\ref{cgaugepterm}) and (\ref{rstterm}) are 
included, but they can still be solved numerically.
Both static and dynamical black hole solutions
were obtained in \cite{Frolov:2006is}. The 
qualitative nature of the solutions depends
on the model parameters, in particular the 
relative size of $\kappa$ and $e^2$.

Let us first consider static black holes. For
$e^2\gg\kappa$ the static solutions are similar to 
the ones with $\kappa=0$ that we have discussed
already. As mentioned above, it is this limit that is
of most interest if we want to model the physics 
of electrons with a large charge-to-mass ratio.
In the opposite limit $\kappa\gg e^2$ the static
solutions are quite different. There is still a 
curvature singularity where $\xi(y)$ goes to zero
inside the black hole but the area function $\psi$
reaches a local minimum before this singularity
is reached. The area bounces back to large values 
and is divergent at the singularity. This means 
that the two-dimensional gravitational coupling
goes to zero at the singularity and, from the
3+1-dimensional point of view, the area of the
transverse two-sphere goes to infinity in a region
of infinite spacetime curvature. This type of ``bounce"
singularity has been seen previously in static 
semiclassical solutions in two-dimensional dilaton 
gravity \cite{Balbinot:1994ee,Birnir:1992by,
Susskind:1992gd} but for physics the more 
interesting question is whether analogous 
behavior occurs in dynamical solutions where
a black hole forms in gravitational collapse.

This brings us to dynamical solutions of the
semiclassical equations with both charged and
neutral matter effects included. This is a more
challenging numerical problem than the one
with only charged matter included, partly because
the equations themselves are more complicated, 
but also due to subtleties involving boundary 
conditions that have to be imposed in the 
strong-coupling region 
\cite{Russo:1992ax,Russo:1992yh}. The numerical
solutions reported in \cite{Frolov:2006is} show 
the formation and subsequent evaporation of a 
black hole with a spacelike singularity. In particular
no bouncing of the area function is observed.
It should be noted, however, that due to redshift
effects a numerical code like the one employed
in \cite{Frolov:2006is}, which is based on a fixed 
coordinate grid, cannot resolve well the 
neighborhood of the event horizon at late stages 
in the evaporation of black hole that starts out
large. This leaves open the possibility that the 
singularity inside a large charged black hole 
has a null portion close to the evaporation 
endpoint \cite{Ori:2006}. Presumably a more
sophisticated numerical approach is needed
to settle this issue.

\begin{acknowledgments}
We thank the organizers of the Workshop on Strings, 
Black Holes, and Quantum Spacetime, Bokwang Phoenix
Park, South Korea, April 21-23, 2006, for the opportunity
to present this work, which was supported in part by 
grants from the Icelandic Research Fund and the 
University of Iceland Research Fund. Section~III is 
based on work in collaboration with Andrei Frolov and 
Kristjan Kristjansson that appeared in 
\cite{Frolov:2005ps,Frolov:2006is}.  We would also like 
to thank Amos Ori for useful discussions.

\end{acknowledgments}

\end{document}